\documentclass[prd,10pt,twocolumn,twoside,a4paper,tightenlines,aps,nofootinbib,preprintnumbers]{revtex4}

\usepackage{amsfonts}
\usepackage{amssymb}
\usepackage[english]{babel}
\usepackage{amsmath}
\usepackage{latexsym}
\usepackage{amscd}
\newcommand{\m}{\mu}
\newcommand{\n}{\nu}
\newcommand{\be}{\begin{equation}}
\newcommand{\ee}{\end{equation}}
\newcommand{\bea}{\begin{eqnarray}}
\newcommand{\eea}{\end{eqnarray}}
\newcommand{\Gam}{\Gamma}
\newcommand{\gym}{\gamma_{ab}}
 
\begin{document}
\title{Type III Einstein-Yang-Mills solutions}

\author{Andrea Fuster${}^{a}$, Jan-Willem van Holten${}^{a}$\\[-2ex] \ }

\affiliation{\mbox{\it ${}^a$ National Institute for Nuclear and High-Energy
Physics
(NIKHEF),}\\ \mbox{\it Kruislaan 409, 1098 SJ, Amsterdam, The Netherlands}}

\date{\today}
\preprint{NIKHEF/2005-009}


\begin{abstract}
We construct two distinct classes of exact type III solutions of the $D=4$ Einstein-Yang-Mills system. The solutions are embeddings of the non-Abelian plane waves in spacetimes in Kundt's class. Reduction of the solutions to type N leads to generalized $pp$ and Kundt waves. The geodesic equations are briefly discussed.
\end{abstract}

\maketitle

\section{Introduction}
Exact solutions of Einstein equations have always attracted much attention. It is somewhat surprising to find exact solutions of such non-linear equations. Many of them were collected in the by now classic book by Kramer et al, which has recently been revised~\cite{Stephani:2003tm}. Among others one finds the non-diverging spacetimes which were studied for the first time by Kundt in~\cite{kundt61}. Solutions which admit a non-expanding and twistfree null congruence are therefore said to belong to the Kundt's class. Such metrics are algebraically special, i.e., of Petrov type III, N, O, II or D. If one restricts the solutions to those with plane wave surfaces they are constrained to be (at most) of type III; they can degenerate to type N and O. Conformally flat solutions include for example the Bertotti-Robinson~\cite{bertotti, robinson} universe and the Edgar-Ludwig metric~\cite{Edgar:1996dd}. An Einstein-Yang-Mills solution of type N was given long time ago in~\cite{Gueven:1979mb} and recovered for $D=4$ supergravity in~\cite{Cariglia:2003ug}. Type III solutions have somehow attracted less attention. Yet, some references considering the type III Einstein-Maxwell system can be found~\cite{hall-carot, cahen-spel}. To our knowledge, the solutions presented in this note are the first type III solutions of Einstein gravity coupled to a Yang-Mills field. \\

We also want to point out that the Kundt's class can be generalized to include a non-zero cosmological constant. Such solutions of type N were first considered in~\cite{pleb-garcd}, followed by ~\cite{Ozsvath:1985qn, Bicak:1999ha, Bicak:1999hb, Podolsky:2002sy}. Type N Einstein-Maxwell solutions were generalized to an arbitrary number of dimensions in~\cite{Obukhov:2003br} and further generalized to Lovelock-Yang-Mills solutions in~\cite{Gleiser:2005tu}. Not very long ago the $\Lambda \neq 0$ generalization of type III spacetimes in~\cite{Stephani:2003tm} was presented in~\cite{Griffiths:2003bk}.

\section{Yang-Mills in Kundt spacetimes}
We consider Kundt's class of metrics, described by the line element: 
\be
\begin{split}
ds^2=&-2du\;(dv+W\;dz+\overline{W}\;d\bar{z}+H\;du)  \\
 &+2P^{-2}dzd\bar{z},\; \;\;P,_{v}=0 \label{metgk}
\end{split}
\ee
Here, $P$ and $H$ are real functions while $W$ is complex; $u=(t-z)/\sqrt{2}$, $v=(t+z)/\sqrt{2}$ are light-cone coordinates and $z=(x+iy)/\sqrt{2}$, $\bar{z}=(x-iy)/\sqrt{2}$ complex conjugate coordinates in the transverse plane. We consider solutions of type III. In this case we can take $P=1$ without loss of generality. As a consequence the Gaussian curvature of the wave surfaces $u$=constant, $K=2P^2(\mbox{ln}\;P),_{z\bar{z}}$, vanishes. Type III solutions are therefore characterized by plane wave surfaces. Details on specific aspects of the metrics (\ref{metgk}) can be found in~\cite{Stephani:2003tm}. \\

It can be seen that two distinct cases arise for vacuum solutions of type III (for references, see~\cite{Stephani:2003tm}). In the first case the function $W$ does not depend on the light-cone coordinate $v$. In the second case the dependence is of the form $W,_{v}=-2/(z+\bar{z})$. These two cases correspond to the Newman-Penrose quantity $\tau$ being zero or $\neq 0$, respectively.  

\subsection{Case $\mathbf{W,_{v}=0}$}

Consider the metric (\ref{metgk}) with:
\begin{subequations}
\bea
P&=&1 \label{1} \\
W&=&W(u,\bar{z}) \label{2} \\
H&=&H^0(u,z,\bar{z}) +\frac{1}{2}\left( W,_{\bar{z}}+\overline{W},_z \right)v \label{3}
\eea
\end{subequations}
Here, $W$ is an arbitrary complex function and $H^0$ is a real function. This is a vacuum solution of Einstein equations if the function $H^0$ satisfies:
\be
\frac{1}{2}R_{33}=H,_{z\bar{z}}^0-\mbox{Re}\left( W,_{\bar{z}}^2+WW,_{\bar{z}\bar{z}}+W,_{\bar{z}u} \right)  
=0  \label{Ruu} 
\ee
Numerical indices are tetrad indices\footnote{Tetrad vectors as in~\cite{Stephani:2003tm}: $\mathbf{e}_1=\mathbf{\bar{e}}_2=\partial_z,\;\;   \mathbf{e}_3=\partial_u+\overline{W}\;\partial_z+W\;\partial_{\bar{z}}-(H+W\overline{W})\;\partial_v,\;\;\mathbf{e}_4=\partial_v$}. All other components of the Ricci tensor vanish. This class of solutions degenerates to type N when $\Psi_3=\frac{1}{2}W,_{\bar{z}\bar{z}}=0$. In this case one can use the remaining coordinate transformation (see~\cite{Stephani:2003tm}) to make $W=0$. The metric reduces then to that of familiar $pp$ waves~\cite{brinkman},
\be  
ds^2 = 2\; dz\;d\bar{z}-2\;du\;dv-H(u,z,\bar{z})\; du^2  \label{pp}  
\ee 
where $H$ satisfies $H,_{z\bar{z}}=0$. \\

The function $W$ can have an arbitrary $u$ dependence but must be at least quadratic in $\bar{z}$ so that $\Psi_3 \neq 0$. This holds for pure radiation solutions as well: they only enter the solution through the function $H^0$ so $W$ is as in vacuum. \\

We can generalize the solution to include a null Yang-Mills field ($\Phi_0=\Phi_1=0$) of the form:
\be
A=\alpha^a(u,z,\bar{z})T_a \;du,\;\; 
{\alpha}^a= \chi^a(u,z)+\bar{\chi}^a(u,\bar{z})  \label{ymfield}
\ee
Here, $\chi^a$ are arbitrary complex functions. This field was considered in curved spacetime for the first time in~\cite{Gueven:1979mb}. The Yang-Mills equation in flat space is just:
\be
\alpha,_{z\bar{z}}=0 \label{ymcomplex}
\ee
This equation remains unchanged in the spacetime (\ref{metgk})-(\ref{3}) because of two reasons. First, the absence of components of the YM field strength of the form $F^{u\m}$ in this geometry. In the second place, the only non-vanishing Christoffel symbols of the form $\Gam_{\rho \n}^{\;\;\;\;\rho}$ are precisely $\Gam_{\rho u}^{\;\;\;\;\rho}$. As a result only the ordinary derivative survives in the (curved) Yang-Mills equation:
\be
\partial_{\lambda}F^{\lambda \m}+\Gam_{\rho u}^{\;\;\;\;\rho}F^{u \m}-[A_u,F^{u \m}]=2\alpha,_{z\bar{z}}=0  
\ee

Of course Eq. (\ref{Ruu}) has to be modified to account for the energy-momentum tensor of the Yang-Mills field: 
\be
T_{uu}=\frac{1}{2 \pi}\gym\;\alpha,_{z}^a\alpha,_{\bar{z}}^b
\ee
Here, $\gym$ is the invariant metric of the Lie group. In tetrad indices we have $T_{33}=T_{uu}$. Therefore, Eq. (\ref{Ruu}) transforms to:
\be
H,_{z\bar{z}}^0-\mbox{Re}\left( W,_{\bar{z}}^2+WW,_{\bar{z}\bar{z}}+W,_{\bar{z}u} \right)=2\gym \;\alpha,_{z}^a\alpha,_{\bar{z}}^b \nonumber 
\ee
It is straightforward to solve for $H^0$:
\be
\begin{split}
H^0(u,z,\bar{z})&=f(u,z)+\bar{f}(u,\bar{z})+2\gym \chi^a \bar{\chi}^b \\  
 &+\mbox{Re}\; \{\left( W,_{u}+WW,_{\bar{z}}    \right)z\}  
\end{split}
\ee
Here $f$ is an arbitrary complex function. Eqs. (\ref{ymfield}) and (\ref{metgk})-(\ref{3}) with an arbitrary (at least quadratic) $W$ and the given $H^0$ describe an exact type III solution of the EYM system. It is a generalized Goldberg-Kerr solution. The type N reduction of this solution has been known for a long time~\cite{Gueven:1979mb}. \\

\subsection{Case $\mathbf{W,_{v}=-2/(z+\bar{z})}$}

We consider now the metric (\ref{metgk}) with:
\begin{subequations}
\begin{gather}
P=1 \label{1v} \\
W=W^0(u,z)-\frac{2v}{z+\bar{z}} \label{2v} \\
H=H^0(u,z,\bar{z}) + \frac{W^0+\overline{W}^0}{z+\bar{z}}v   
 -\frac{v^2}{(z+\bar{z})^2}  \label{3v} 
\end{gather}
\end{subequations}
Here, $W^0$ is an arbitrary complex function and $H^0$ is a real function. This is a vacuum solution of Einstein equations if the function $H^0$ satisfies the differential equation:
\be
\frac{1}{2}R_{33}^0=(z+\bar{z})\left( \frac{H^0+\overline{W}^0W^0}{z+\bar{z}} \right)_{,_{z\bar{z}}}-W,_{z}^0\overline{W},_{\bar{z}}^0 
=0  \label{eqdif}
\ee
This is the only non-vanishing component of the Ricci tensor\footnote{The superscript $^0$ in $R_{33}$ denotes the $v$-independent part of $R_{33}$.}. This class of solutions degenerate to type N as well when $\Psi_3=\overline{W},_{\bar{z}}^0/(z+\bar{z})=0$. In this case one can transform $W^0$ to zero~\cite{Stephani:2003tm} and the resulting metric is known as Kundt waves. \\

We consider the field (\ref{ymfield}) in this background. Features in this geometry are the same ones as in the previous case, except for an extra term in the Yang-Mills equation:
\be
(\partial_z +\Gam_{uz}^{\;\;\;\;u} +\Gam_{vz}^{\;\;\;\;v})F^{zv} +(z \leftrightarrow \bar{z})=0
\ee
After further inspection this extra term is seen to vanish as $\Gam_{uz}^{\;\;\;\;u}=-\Gam_{vz}^{\;\;\;\;v}$, $\Gam_{u\bar{z}}^{\;\;\;\;u}=-\Gam_{v\bar{z}}^{\;\;\;\;v}$. In this way, the Yang-Mills equation in the spacetime (\ref{metgk}), (\ref{2v}), (\ref{3v}) is still the same as in flat space, Eq. (\ref{ymcomplex}). Finally, Eq. (\ref{eqdif}) has to be modified to include the YM energy-momentum tensor $T_{33}=(\gym/2\pi)\;\alpha,_{z}^a\alpha,_{\bar{z}}^b$. \\

We refine our ansatz in order to solve explicitly for $H^0$. We choose the specific function $W^0=\;g(u)z$, $g(u)$ being an arbitrary complex function. This is the simplest form of $W^0$ such that $\Psi_3$ is non-vanishing. On the other hand we limit the Yang-Mills field (\ref{ymfield}) to:
\be
\chi^a=\lambda^a(u)\;z \label{ebound}
\ee
Here $\lambda^a(u)$ are bounded complex functions. This field was considered in Minkowski space for the first time in~\cite{Coleman:1977ps} and is referred to as non-Abelian plane waves\footnote{The original field in~\cite{Coleman:1977ps} was $\chi^a=\lambda^a(u)\;z+\mbox{some function of}\;u$ but the later term can be gauged away without affecting gauge-invariant quantities, as pointed out by Coleman.}. The field (\ref{ebound}) is the only form of (\ref{ymfield}) which gives rise to a bounded energy-momentum tensor. Under these assumptions the equation to solve reads:
\be
(z+\bar{z})\left( \frac{H^0+g\bar{g}z\bar{z}}{z+\bar{z}} \right)_{,_{z\bar{z}}}-g(u)\bar{g}(u)=2\gym\lambda^a(u)\bar{\lambda}^b(u) \nonumber 
\ee
It is not difficult to see that:
\be
\begin{split}
H^0&= (f(u,z)+\bar{f}(u,\bar{z}))\;(z+\bar{z})-g\bar{g}z\bar{z}  \\
&+\sigma(u)(z+\bar{z})^2\left\{ \mbox{ln}(z+\bar{z})-1 \right)  
\end{split} \label{Kdif}
\ee
Here $f$ is again an arbitrary complex function and $\sigma$ is the real function $\sigma(u)=2\gym\lambda^a(u)\bar{\lambda}^b(u)+g(u)\bar{g}(u)$. Eqs. (\ref{ymfield}), (\ref{ebound}) and (\ref{metgk}), (\ref{1v})-(\ref{3v}) with the given $W^0$, $H^0$ describe another exact type III solution of the EYM system. The corresponding vacuum solution ($\lambda^a\equiv0$) has to our knowledge not been considered before. The case $g(u)\equiv0$ yields the type N reduction and can be regarded as generalized Kundt waves. A related type N solution was very recently given in~\cite{Gleiser:2005tu}, for a Yang-Mills field of arbitrary higher (polynomial) dependence on $z$, $\bar{z}$; our (type N) solution is obtained by translating the line element into the Kundt canonical form and considering $N=0$, $D=4$.

\subsection{Geodesics}
For completeness, we present the geodesic equations corresponding to the two classes of solutions found, without attempting to provide detailed solutions here. \\  

$\cdot$ Case $W,_{v}=0$ \\

In order to get a feeling for the behavior of geodesics we choose the specific $W=\bar{z}^2$. This is the simplest $W$ such that the solution presented in subsection {\bf{A}} is of type III. The geodesics are described in terms of real spatial coordinates $x$, $y$ by: 
\begin{gather}
\ddot{u}-\sqrt{2}x\dot{u}^2=0 \label{geod1} \\
\ddot{x}+\dot{u}^2 \left(H,_{x}-x(x^2-y^2) \right)+ 2\sqrt{2}y\;\dot{u}\dot{y}=0 \label{geod2} \\
\ddot{y}+\dot{u}^2 \left( H,_{y}-2yx^2 \right)-2\sqrt{2}y\;\dot{u}\dot{x} =0 \label{geod3} \\ 
\dot{x}^2+\dot{y}^2-2\dot{u}\dot{v}-2H\dot{u}^2-\sqrt{2}(x^2-y^2)\dot{x}\dot{u}  -2\sqrt{2}yx\;\dot{y}\dot{u}=\epsilon \label{geod4} 
\end{gather}
Here, the overdot denotes a proper time derivative and $H$ is: 
\be
H=f+\bar{f}+ 2(x^4-y^4)+ 2\gym\chi^a \bar{\chi}^b+\sqrt{2}xv  \label{HWquad}
\ee
We have considered the normalization condition for the $4$-velocity, Eq. (\ref{geod4}), instead of the (complicated) equation for $v$. There, $\epsilon=-1,0$ for timelike and null geodesics, respectively. \\

It can be advantageous to use $u$ as affine parameter instead of the proper time. If we assume $\dot{u}>0$, from Eq. (\ref{geod1}): 
\be
x=\frac{1}{\sqrt{2}}\frac{d\;(\mbox{ln} \dot{u})}{du} \label{xfunujw}
\ee
Now, in the hyperplane $y=0$, by inserting Eqs. (\ref{xfunujw}) in (\ref{geod2}), (\ref{geod4}) and recalling $\dot{x}=\dot{u}\;(dx/du)$, a coupled system for $x(u)$, $v(u)$ can be obtained. \\

On the other hand, in the hyperplane $x=0$ Eqs. (\ref{geod1}) and (\ref{geod3}) simplify to

\bea
\ddot{u}&=&0 \label{geod1x0} \leftrightarrow \dot{u}=\mbox{constant}\equiv \gamma \\
\ddot{y}+\dot{u}^2  H,_{y} &=&0 \label{geod3x0} 
\eea
taking the same form as in the type N reduction (see for example~\cite{vanHolten:1999jc}). Analytical differences come from the quartic term of (\ref{HWquad}) which is absent in the type N. This difference can be physically illustrated when we consider the functions $f=\kappa(u)\;z^2$, with $\kappa(u)$ an arbitrary real function\footnote{This choice is well motivated as it corresponds to homogeneous $pp$ waves in the (vacuum) type N reduction.}, and $\chi^a=\lambda^a(u)\;z$ in Eq. (\ref{HWquad}). Then, the motion in the $y$ direction is governed by:
\be
d^2 y/du^2-8y^3+2(\gym\lambda^a(u)\bar{\lambda}^b(u)-\kappa(u))y=0 \label{nonlosc}
\ee
This equation describes a (parametric) non-linear oscillator in a potential $V(y)=y^2(\gym\lambda^a\bar{\lambda}^b-\kappa-2y^2)$. In the type N reduction there is no cubic term in Eq. (\ref{nonlosc}) and the motion reduces to that of an harmonic oscillator. \\

Non-linear oscillators can have chaotic motion. Geodesic chaotic motion has been found in the related case of non-homogeneous vacuum $pp$ waves ($f\sim z^n,\;n>2$), see~\cite{Podolsky:1998ez}. However, chaotic behavior is not always present in non-linear systems and therefore possible chaotic motion arising from Eq. (\ref{nonlosc}) would need further investigation. \\

$\cdot$ Case $W,_{v}=-2/(z+\bar{z})$ \\

We take the $u$ dependence out of $W^0$ respect to the function considered in subsection {\bf{B}}, so now $W^0=z$. Geodesics in this second case have the more complicated form:
\begin{gather}
\ddot{u}-(1-v/x^2)\dot{u}^2-2/x\;\dot{u}\dot{x}=0 \\
\begin{align}
\ddot{x}+\dot{u}^2 \left(H,_{x}+(x-2v/x)(1-v/x^2) \right)&  + 2\left(1-2v/x^2\right)\dot{u}\dot{x} \nonumber \\
&-2/x\;\dot{v}\dot{u}=0 
\end{align} \\
\ddot{y}+\dot{u}^2 \left( H,_{y}+y (1-v/x^2)\right)+2y/x\;\dot{u}\dot{x} =0  \\
\dot{x}^2+\dot{y}^2-2\dot{u}\dot{v}-2H\dot{u}^2-2(-2v/x +x)\;\dot{x}\dot{u} +2y\;\dot{y}\dot{u}=\epsilon 
\end{gather}

Again, the normalization condition of the $4$-velocity takes the place of the equation for $v$. The function $H$ is
\be
H=H^0+v-\frac{v^2}{2x^2}
\ee
and $H^0$ is given by (\ref{Kdif}) but with $g(u)\equiv 1$. \\

The geodesic equations can again be rewritten
as differential equations in terms of an affine variable $u$
instead of the proper time. In particular, for the first equation:
\be
\frac{d}{du} \ln \left( \frac{\dot{u}}{x^2} \right) = 1 - \frac{v}{x^2}
\ee
Then the remaining geodesics become equations for
$(x,y,v)$ in terms of $u$. \\

The existence of a singularity at $x=0$ is clear from the geodesic equations, but not its character. Geodesics for a general vacuum type N or conformally flat pure radiation solutions have been analyzed in~\cite{Podolsky:2004qu}, where it was concluded that $x=0$ is a physical singularity. Whether this is also the case for our type III solutions remains an open question.  \\

\section{Conclusions}
In~\cite{Tafel:1986tm} Tafel claimed that null twistfree solutions of the Yang-Mills equations were exhausted by the type N (class of) solutions given in~\cite{Gueven:1979mb}. A new type N solution was recently given in~\cite{Gleiser:2005tu}. We have shown that more general solutions exist, as the two type III (classes of) EYM solutions presented here are also null and twistfree. \\

It has been proved~\cite{Pravda:2002us} that all spacetimes with vanishing curvature invariants of all orders: $1$) are Kundt metrics of Petrov type III, N or O, and $2$) are of Pleba\'{n}ski-Petrov (PP) type N or O (null radiation or vacuum). These results have been generalized to higher dimensions in~\cite{Coley:2004hu}. Such spacetimes have been called VSI spacetimes. Our solutions clearly meet requirements $1$) and $2$) and are therefore new examples of $D=4$ VSI spacetimes. As such, they might have interesting physical applications in supergravity and string theory, in the fashion of $pp$ waves. \\

\begin{acknowledgments}
A.F. would like to thank N. Van den Bergh for discussions. J.W.v.H.\ whishes to thank the DESY theory division, where part of this work was performed, for its hospitality. This work is part of the programme FP52 of the Foundation for 
Research of Matter (FOM). 
\end{acknowledgments}

\bibliography{noterv}

\end{document}